\documentclass[showpacs,preprintnumbers,a4paper,aps,12pt]{revtex4}

\usepackage{epsfig}

\newcommand{\be}{\begin{equation}}
\newcommand{\ee}{\end{equation}}
\newcommand{\bd}{\begin{displaymath}}
\newcommand{\ed}{\end{displaymath}}
\newcommand{\bea}{\begin{eqnarray}}
\newcommand{\eea}{\end{eqnarray}}

\begin{document}

\title{ Graphene Multi-Protonation: a Cooperative Mechanism for Proton Permeation  }

\author{Massimiliano Bartolomei \footnote{Corresponding author,
    e-mail:maxbart@iff.csic.es}}
\affiliation{Instituto de F\'{\i}sica Fundamental, Consejo Superior de 
Investigaciones Cient\'{\i}ficas (IFF-CSIC), Serrano 123, 28006 Madrid, Spain}

\author{Marta I. Hern{\'a}ndez}
\affiliation{Instituto de F\'{\i}sica Fundamental, Consejo Superior de 
Investigaciones Cient\'{\i}ficas (IFF-CSIC), Serrano 123, 28006 Madrid, Spain}

\author{Jos{\'e} Campos-Mart{\'\i}nez}
\affiliation{Instituto de F\'{\i}sica Fundamental, Consejo Superior de 
Investigaciones Cient\'{\i}ficas (IFF-CSIC), Serrano 123, 28006 Madrid, Spain}

\author{ Ram{\'o}n Hern{\'a}ndez Lamoneda}
\affiliation{Centro de Investigaciones Qu{\'\i}micas, Universidad Aut\'onoma del  Estado
de Morelos, 62210 Cuernavaca, Mor. M\'exico}

\date{\today}


\begin{abstract}

 The interaction between protons and graphene is attracting a large
interest due to recent experiments showing that these charged species permeate
through the 2D material following a low barrier ($\sim$ 0.8 eV) activated process.
A possible explanation involves the flipping of a chemisorbed proton
(rotation of the C-H$^+$ bond from one to the other side of the carbon layer)
and previous studies have found so far that the energy barriers (around 3.5 eV)
are too high to explain the experimental findings. Contrarily to the
previously adopted model assuming an isolated proton, in this work we
consider protonated graphene at high local coverage and explore the role played by nearby
chemisorbed protons in the permeation process. By means of density
functional theory calculations exploiting large molecular prototypes for graphene
it is found that, when various protons are adsorbed on the same carbon hexagonal
ring, the permeation barrier can be reduced down to 1.0 eV. 
The related mechanism is described in detail and could shed a new light 
on the interpretation of the experimental observations for proton permeation 
through graphene.
  
\end{abstract}


\maketitle


\newpage

\section{Introduction}

This work is partly motivated by recent experimental work on the permeation at room 
temperature of protons through graphene\cite{Hu-2014}. In that study, Geim 
and coworkers reported conductance and mass spectroscopy measurements of 
proton transport through pristine graphene and found that,
in a temperature range of 270-330 K, the process exhibits an Arrhenius-type 
behavior with a rather low activation energy (about 0.8 eV).
Moreover, protons permeate through this two-dimensional crystal about ten
times faster than deuterons\cite{Lozada-2016}. These discoveries were 
absolutely unexpected since graphene was believed to be completely impermeable 
to all atoms and molecules under ambient conditions\cite{Berry-2013,Miao-2013}.
A number of works have subsequently appeared -both 
experimental\cite{Achtyl-2015,Walker-2015,LiuXing-2017,ZhangS-2017,
Bukola-JACS-2018,ZhangSheng2018} and
theoretical\cite{Kroes-2017,XuAo-2017,Ekanayake-2017,Feng-2017,
Poltavsky-2018,Mazzuca-2018}- not only stimulated by the promise of important 
applications in hydrogen technology but also with the aim to uncover the 
microscopic mechanisms underlying these observations.

 Despite much recent progress on the theoretical insight into the
  processes leading to such a facile permeation of protons
  through graphene \cite{Kroes-2017,XuAo-2017,Ekanayake-2017,Feng-2017,
    Poltavsky-2018,Mazzuca-2018}, we consider that a complete and satisfactory
  understanding has not been achieved yet (e.g. see
  Refs.\cite{Kroes-2017,XuAo-2017} for some discussions). 
The possible role of surface defects in the transport
process\cite{Achtyl-2015,Walker-2015} has been also indicated.
 Some works\cite{Poltavsky-2018,Mazzuca-2018} have emphasized the role of 
quantum tunneling in effectively lowering the energy barrier and on isotope 
selectivity, using models that assume that protons/deuterons are free 
particles. In the experiments, however, protons are initially moving within 
an aqueous medium (hydrated Nafion or HCl solution), so other
works\cite{XuAo-2017,Ekanayake-2017,Feng-2017}  have more realistically
considered protons to be bound 
to water (as H$_3$O$^+$) in the reactants states.  
As Shi et al\cite{XuAo-2017} indicate, two possible modes for the penetration of a
hydrated proton can be distinguished. The first one (dissociation-penetration) involves
the removal of the proton from the water network as it crosses the graphene membrane
through the hollow of a carbon ring. It has been found that this mechanism is
unlikely due to the large proton affinity of water\cite{XuAo-2017,Ekanayake-2017}.
Interestingly, Feng {\em et al}\cite{Feng-2017} have recently found, by means of
first principles calculations, that the proton transport can be largely facilitated
if various carbon atoms of the graphene layer (close to the crossing region) are
in sp$^3$ configurations due to hydrogenation. In the second mechanism
(adsorption-penetration), a proton is transferred from the aqueous media to the graphene
surface, where it becomes chemisorbed at the top of a carbon atom
and, in a second step, it flips through the hollow from the original chemisorbed
site to a related one on the other side of the layer. Previous works\cite{Feng-2017} indicate
that the barrier for proton transfer from water to a chemisorbed state is small and therefore
the first step appears to be feasible at ambient conditions. However, the chemisorbed state is
very stable, hence the barrier for the proton flipping becomes 
extremely high ($\sim$ 3.5 eV\cite{Miao-2013,Kroes-2017,Feng-2017}). 

In this work, we assume that the chemisorbed proton is not initially isolated
but surrounded by a number of protons that are chemisorbed as well and, in this way,
we aim to explore the role of these neighbors on the flipping process.
In other words, it is supposed that protons have been already transferred
from the water network to a set of chemisorption sites so that the graphene sheet becomes
partially protonated on one of its sides. It is found that, if two protons respectively 
attach to two consecutive carbon atoms, the permeation mechanism is completely different to
that of an isolated chemisorbed proton. First, the energy barrier drops quite significantly
(down to about 1.0 eV). Second, in contrast with the isolated case, where the reaction path goes
through a planar transition state near the center of the carbon ring, the path for the
multiprotonated case involves the insertion of the flipping proton into the middle of 
an effectively broken C-C bond. This bond is restored once the flipping process comes
to an end, therefore preserving the stability of the carbon layer.
These findings can help to rationalize 
  the proton permeation observations\cite{Hu-2014}; additionally, they could provide
  some clues about properties of hydrogenated graphene\cite{Balog-2010,Martinazzo-2018} or 
about astrochemical processes involving hydrogen/protons and
carbonaceous surfaces\cite{Gago-Nat-2014,Gago-jpcc-2014}.

The mechanism here investigated bears some resemblance with that proposed
  in a study by Lee {\em et al}\cite{Lee-JACS-2001} on the electrochemical storage
of hydrogen within a narrow single-walled carbon nanotube (SWCNT), where a low energy
reaction path was identified for a hydrogen atom flipping from the external to the internal
side of the nanotube. 
 In addition, it is also worth noting that  our study apparently shows
 similarities with that of  Feng {\em et al}\cite{Feng-2017}, in the sense
 that in both cases it is concluded that the barrier for proton penetration notably
 decreases upon local protonation/hydrogenation. However, the involved microscopic
 mechanisms are completely different.
A detailed comparison with the above mentioned works is provided at the end of Section 3.

The paper is organized as follows. Computational methods are described in the following section.
Next, results are presented and discussed, starting with an analysis of the stability of 
protonated graphene and continuing with a study of the permeation process as a function
of the number of chemisorbed protons along a carbon ring.  The report closes
with a summary where further lines for research are indicated.

\section{Computational Methods}
\label{sec.2}

We have carried out electronic structure calculations to study the role played
by an increasing number of protons
($n=1-6$), all chemisorbed along a given carbon ring of graphene, in their
penetration through the 2D layer, which have been described exploiting a
molecular model. 
In previous works the coronene molecule was found to be a sufficiently accurate model
for the study of the
physical adsorption of single atoms\cite{grapheneours:2013,coron-h2:2017} or the sticking of
hydrogen atoms\cite{Morokuma-2012,Michaelides-2014} to graphene.
Here, a larger molecular prototype is required to correctly describe a more
complex process involving a larger number of adsorbed species and, to
this end, we have found that the use
of circumcoronene (C$_{54}$H$_{18}$) is sufficiently adequate for the present purposes. A significant
advantage of using molecular prototypes is that an arbitrary number of protons
can be included in
the calculation, whereas periodic calculations suffer from the problem of having to neutralize the
unit cell to converge and this procedure becomes less reliable when increasing the number of charges.

DFT calculations have been performed for the optimization of the protonated circumcoronene structures
by using the PBE\cite{pbe:96} functional together with the cc-pVTZ\cite{Dunning} basis set. These
calculations  were found to be in good agreement with benchmark MP2/aug-cc-pVTZ computations carried
out for a smaller carbon plane  prototype such as coronene.
 Additional calculations involving larger prototypes
(C$_{96}$H$_{24}$ and C$_{150}$H$_{30}$) were carried  out with a reduced
6-311+G\cite{Pople:80} basis set.  We have verified
that, in the case of circumcoronene, this smaller set provides energy variations that are in good agreement
(within few percents) with those obtained with the cc-pVTZ\cite{Dunning} basis set.
All reported energies correspond to stationary points whose correct nature has
been verified by carrying out harmonic frequency calculations, used in turn to
estimate zero-point energy and thermal corrections (at 298 K and 1 atm) to
thermodynamic properties. The enthalpy of the proton in the gas phase have
been estimated to be $\frac{5}{2}RT$ as a result of the application of the
standard statistical mechanics and ideal gas expressions. 
Intrinsic reaction coordinate calculations have been
employed to check that reactants and products are indeed connected with the
transition states for various of the permeation processes.

All DFT computations have been performed by using the Gaussian 09 code\cite{g09}.

\section{Results and Discussion}

\begin{figure}
\includegraphics[width=6.cm,angle=0.]{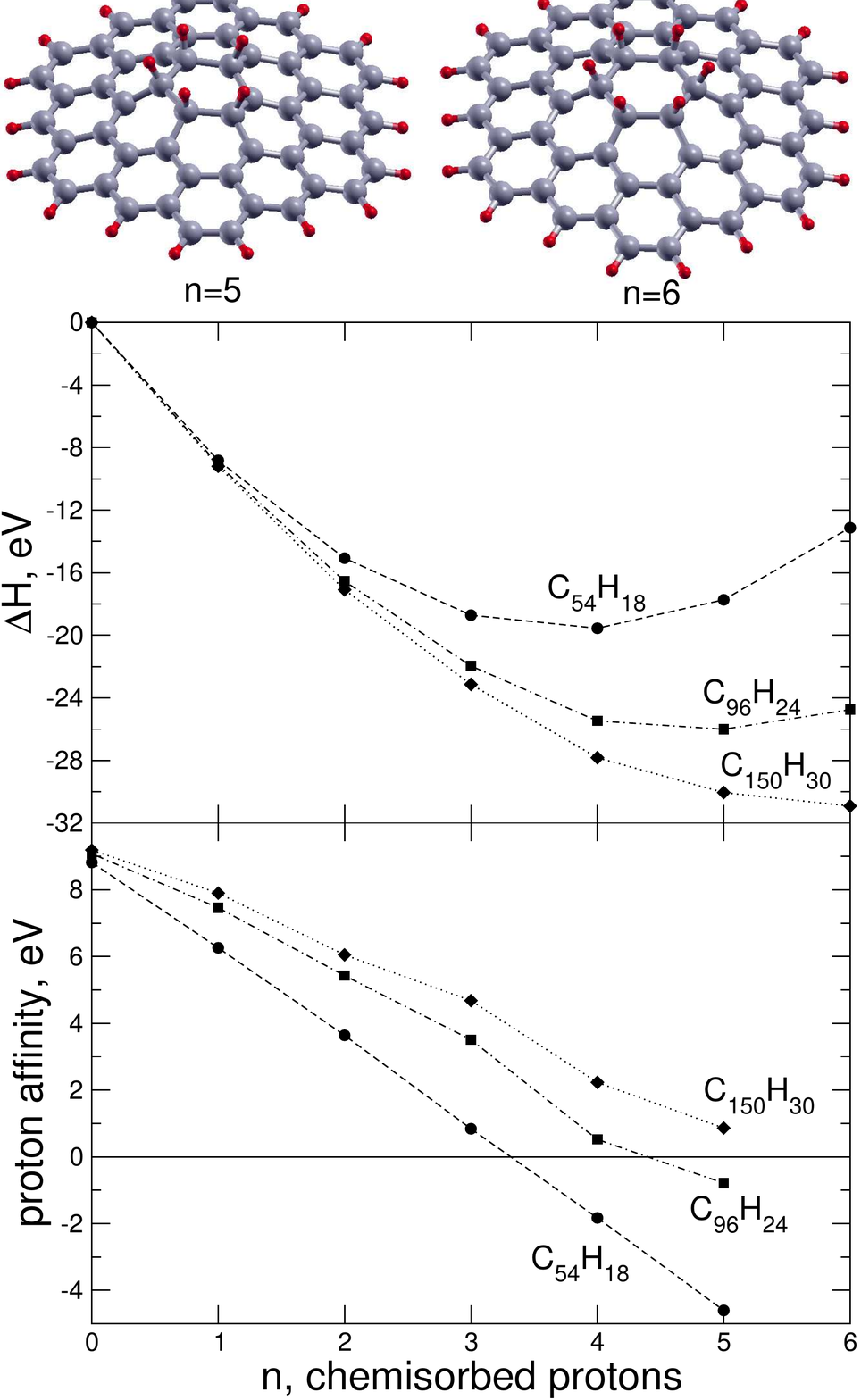}
\caption[]{ Sequential protonation of a graphenic single ring.
  Upper panel: corresponding enthalpy variation with respect to the unprotonated 
  graphene molecular prototype ($n$=0) and isolated protons. 
  Lower panel: proton affinity of each protonated graphene prototype ((C$_x$ H$_{y+n}$)$^{n+}$, n=0-5).\\
The dashed, solid and dotted lines  correspond to the
 C$_{54}$H$_{18}$  (circumcoronene), C$_{96}$H$_{24}$ (circumcircumcoronene)
 and C$_{150}$H$_{30}$ (circumcircumcircumcoronene)   prototypes, respectively.}
\label{fig1}
\end{figure}

First, we address the question of graphene affinity for the chemical adsorption of protons.
Structures of circumcoronene for the sequential sticking of protons above its central ring 
are illustrated in Fig.~\ref{fig1}. In the upper panel of Fig.~\ref{fig1},
we report the enthalpy variation ($\Delta H$) as a function of the number $n$
of chemisorbed protons, which  corresponds to the gas phase process 

\begin{equation}
C_x H_y + nH^+  \rightarrow (C_x H_{y+n})^{n+},
\end{equation}

\noindent
where $C_x H_y$ represents the unprotonated ($n$=0)  graphene prototype.
In the lower panel we depict instead the proton affinity of each
protonated graphene prototype, which is defined as 
 -$\Delta H$ of the following process

\begin{equation}
(C_x H_{y+n})^{n+}  + H^+  \rightarrow (C_x H_{y+n+1})^{(n+1)+},  n=0-5.
\end{equation}

\noindent
It can be seen that in the case of circumcoronene (C$_{54}$H$_{18}$)  the consecutive addition of
a proton to the inner ring is an energetically favourable process: in fact,
positive proton affinities are obtained for $n$ up to 3, that is up to four
protons could be adsorbed. A similar result was previously theoretically
described\cite{benzene_proton:02} for the multiprotonation of benzene, for which the addition
of up to 3 protons was found to lead to stable molecular structures, that is
preserving the hexagonal ring.
 In this case, it seems also clear that the stability of
the most protonated ($n$=5,6) fragments depends on the size of the considered graphene flakes. 
In fact, when larger graphene molecular prototypes (C$_{96}$H$_{24}$ and
C$_{150}$H$_{30}$) are taken into account we observe (upper panel of Fig.~\ref{fig1})  progressively larger
enthalpy variations which seem to tend towards a converged profile as a
function of $n$.  
In particular, in the case of the circumcircumcircumcoronene  (C$_{150}$H$_{30}$) prototypes, positive proton
affinities are found for all considered protonated fragments, that is the
chemisorption of up to 6 protons is feasible and leads to stable molecular structures. 
These results suggest that the saturation of a graphenic ring with protons is
energetically possible in the gas phase and we can expect even more favorable proton
affinities for larger graphene flakes.

 It is also crucial for the adopted model to mention that
 the local properties of the multiprotonated site are found to barely depend
  on the size of the graphene flake. For example, partial charges
  and bond distances of the central ring of a 6 times protonated circumcoronene are very
  similar to those of an analogously protonated circumcircumcoronene. 
Note that the net charge
  of this central ring is about +0.5 so that an excess charge of about +5.5 
  spreads over the rest of the flake. It is reasonable that this excess
  charge becomes more easily distributed as the size of the prototype increases, hence 
  this feature must be at the origin of the larger proton affinities of
  the bigger prototypes (Fig.~\ref{fig1}). In addition, we have also noticed that, 
  once the C-H$^+$ bond is formed, most of the proton character is lost as the partial
  charge on hydrogen is very close (slightly larger)  to that typical of an usual C-H bond. 
  Nevertheless, to remind the reader that the whole graphene flakes have the charge
  of the added protons, the positive charge on the hydrogen atoms will be retained
  in the notations used below.

\begin{figure}
\includegraphics[width=7.5cm,angle=0.]{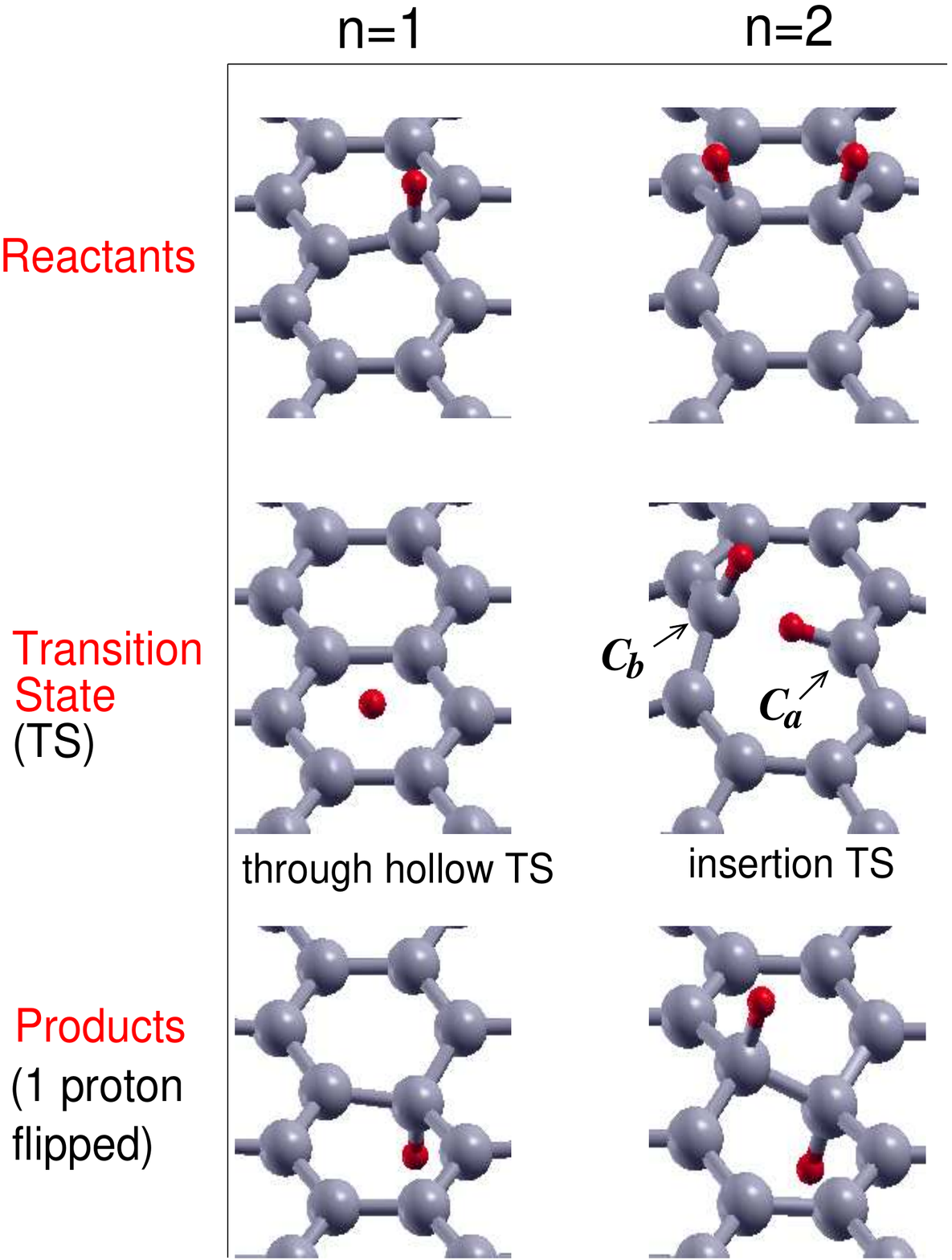}
\caption[]{ Most favorable proton flipping mechanisms for the consecutive
  protonation (up to two protons) of a graphenic single ring. 
 The first row shows the reactants configuration with adjacent protons
 chemisorbed on the same side of the carbon plane.
 The second row shows the transition state (TS) configuration. For $n$=2, carbon atoms bound 
 to the flipping proton and to the adjacent proton are
 indicated as C$_a$ and C$_b$, respectively.
 The last row shows the products configuration with one proton flipped
 on the other side of the carbon plane. 
 The corresponding electronic energy and enthalpy balances are reported in Table \ref{tab:1}.}
\label{fig2}
\end{figure}

\begin{figure}
\includegraphics[width=7.5cm,angle=0.]{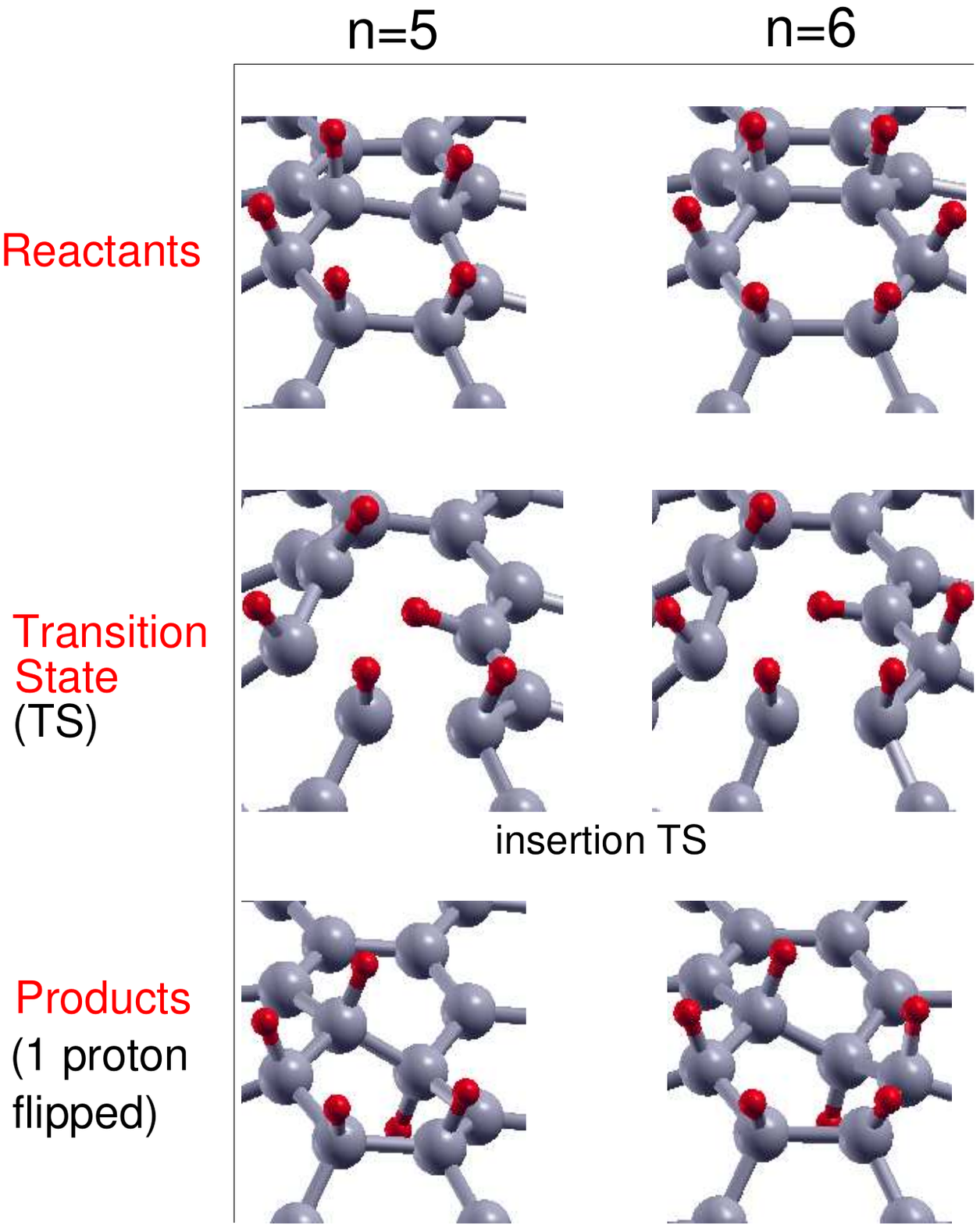}
\caption[]{As in Fig. \ref{fig2} but with five and six protons chemisorbed on the
  graphenic single ring.}
\label{fig3}
\end{figure}

Having established the stability of the graphene prototypes when several protons are
chemisorbed on a benzenic ring, we start studying the permeation process for one and two
chemisorbed protons as well as the underlying microscopic mechanism. 
 In Fig.~\ref{fig2}, the proton penetration from one side to the other of the 
carbon plane is considered for the case of a single chemisorbed proton ($n=1$)
as well as for that of two protons attached to two consecutive carbon atoms ($n=2$). For the isolated proton 
the transition state (TS) corresponds to a planar structure: the C-H$^+$ bond rotates
to locate the proton near the center of the ring (termed as ``hollow'' TS) and continues the rotation to end
up in an equivalent chemisorption state at the other side. Notice that at the TS the original
C-H$^+$ bond has been broken.  However, as pointed out by Miao and
  coworkers\cite{Miao-2013}, there are chemical interactions of the proton with
  the surrounding carbon atoms in the ring.
 Still, the corresponding activation energy ($\Delta E_a$) is quite high and close to 3.5 eV as
reported in Table~\ref{tab:1}. Similar results have already been obtained from
previous calculations at the DFT level\cite{Miao-2013,Kroes-2017,Feng-2017}. 
However, for two nearby chemisorbed protons the most favourable reaction path is quite different:
the related TS is not planar and we observe the insertion of the proton
through the middle of the C-C bond connecting to the closest chemisorbed additional
proton (``insertion'' TS). For this to occur, the length of that C-C bond changes from 1.55
(reactants) to
2.39 \AA \, (TS), so the bond effectively breaks. Interestingly, this breaking of the C-C bond
does not provoke a large energetic penalty as could be expected. On the contrary,
as shown in Table~\ref{tab:1}, the corresponding activation energy is
about 2.8 eV, lower than that of the single chemisorbed proton. In part this is due to the
fact that, during the insertion, the C-H$^+$ bond is preserved as opposed to the $n=1$
case where such a bond is broken.

 \begin{table}
 \caption{
Electronic energy and enthalphy variations associated to the most favorable
single proton flipping process for an increasing number ($n$) of adjacent chemisorbed protons
 on a graphenic carbon ring. The cases of $n$=1,2,5 and 6 are illustrated in
 Figs.~\ref{fig2} and \ref{fig3}. The first two columns show the energy balances
 between reactants (R) and transition state (TS) while the last columns, those
 between reactants and products (P). The last line reports the values for the
 arrangement  of five protons on a ring plus one proton on a surrounding ring
 (Fig.~\ref{fig5}), leading to the lowest activation energy. Values are in eV.
  \label{tab:1}}
\setlength{\tabcolsep}{12pt}
 \begin{tabular}{ccccc}
  &\multicolumn{2}{c}{R$\rightarrow$TS} & \multicolumn{2}{c}{R$\rightarrow$P}    \\
 \hline
 chemisorbed  & $\Delta E_a$  &  $\Delta H_a$  & $\Delta E_r$  & $\Delta H_r$  \\
 protons &   &    &   &   \\
\hline
 n=1    & 3.44 & 3.24 & 0.00 & 0.00 \\
 n=2 & 2.77  & 2.65 & -0.39  & -0.39 \\
 n=3 & 2.29  & 2.17  & -0.38 & -0.42  \\
 n=4 &  1.76   &  1.69  & -0.37   &  -0.37   \\
 n=5  &  1.53   &  1.44  & -0.46   & -0.47  \\
 n=6  &  1.61   &  1.50 &  -1.21   & -1.20  \\
\hline 
 n=5+1  &  1.01   &  0.95 &  -0.55   & -0.56  \\
\end{tabular}
\end{table}

To get insight into this new mechanism, we have examined in detail the
structures of the corresponding stationary points (right panel of Fig.~\ref{fig2}).
Let us denote H$^+_a$ and H$^+_b$ as the flipping and adjacent protons, respectively,
and C$_a$ and C$_b$, as the carbon atoms linked to those protons (see 
right central panel of Fig.~\ref{fig2}). Initially, these carbon atoms 
exhibit four bonds 
in a sp$^3$-like hybridization (i.e., C$_a$ is bound to H$^+_a$, to C$_b$ and to
two other adjacent carbon atoms of the graphenic network, the corresponding C-C distances,
1.50-1.55 \AA, being typical of single bonds). At the TS, the C$_a$-C$_b$ bond is broken but 
the C$_a$ atom keeps its bonding with the three remaining atoms, showing, however,
quite different bond distances and angles (C-C distances are reduced to 1.36 \AA).
Indeed, these four atoms nearly exhibit a planar geometry (with 
$\widehat{{\rm C-C}_a{\rm -H} ^+_a}$ and $\widehat{{\rm C-C}_a{\rm -C}}$ angles
of 113$^o$ and 134$^o$, respectively), indicating that
the C$_a$ atom approximately adopts a sp$^2$ configuration. Analogously, the three
remaining bonds around the C$_b$ atom have transformed towards a trigonal planar geometry
(angles $\widehat{{\rm C-C}_b{\rm -H} ^+_b}$ and $\widehat{{\rm C-C}_b{\rm -C}}$
 being 118.5$^o$ and 122$^o$, respectively), thus showing also C$_b$ a sp$^2$ configuration.
Although the distance between the flipping proton and the opposite carbon atom at the TS,
H$^+_a$-C$_b$, is short (1.32 \AA) it does not correspond to a bonding interaction but rather
to a repulsive one. This is to be expected considering that, as stated before,
the proton character in the C$_a$H$^+_a$ bond has been mostly lost and therefore it can only experience
a steric repulsion with the C$_b$ atom. This will be further confirmed below as a correlation between
the H$^+_a$-C$_b$ distances and the barrier heights.  Given that a C-C
  bond has been broken it is natural to enquire whether an open-shell
  structure with unpaired electrons characterizes the TS. We have performed
  stability tests of the closed-shell PBE determinant which prove it is stable
  and corresponding to the lowest energy solution (from the comparison 
with triplet states and unrestricted solutions). Therefore, as a qualitative explanation for the relatively low activation energy of this
insertion process, it can be said that the cost of breaking the C-C bond at the TS is compensated
by the transformation of the remaining bonds to stronger sp$^2$-type ones.
We have checked that, when additional adsorbed protons are added to the same ring of carbon atoms,
the penetration process most favourably occurs through the same kind of insertion TS, as can be seen,
for instance, in Fig.~\ref{fig3} for the case of five and six protons.

\begin{figure}
\includegraphics[width=7.5cm,angle=0.]{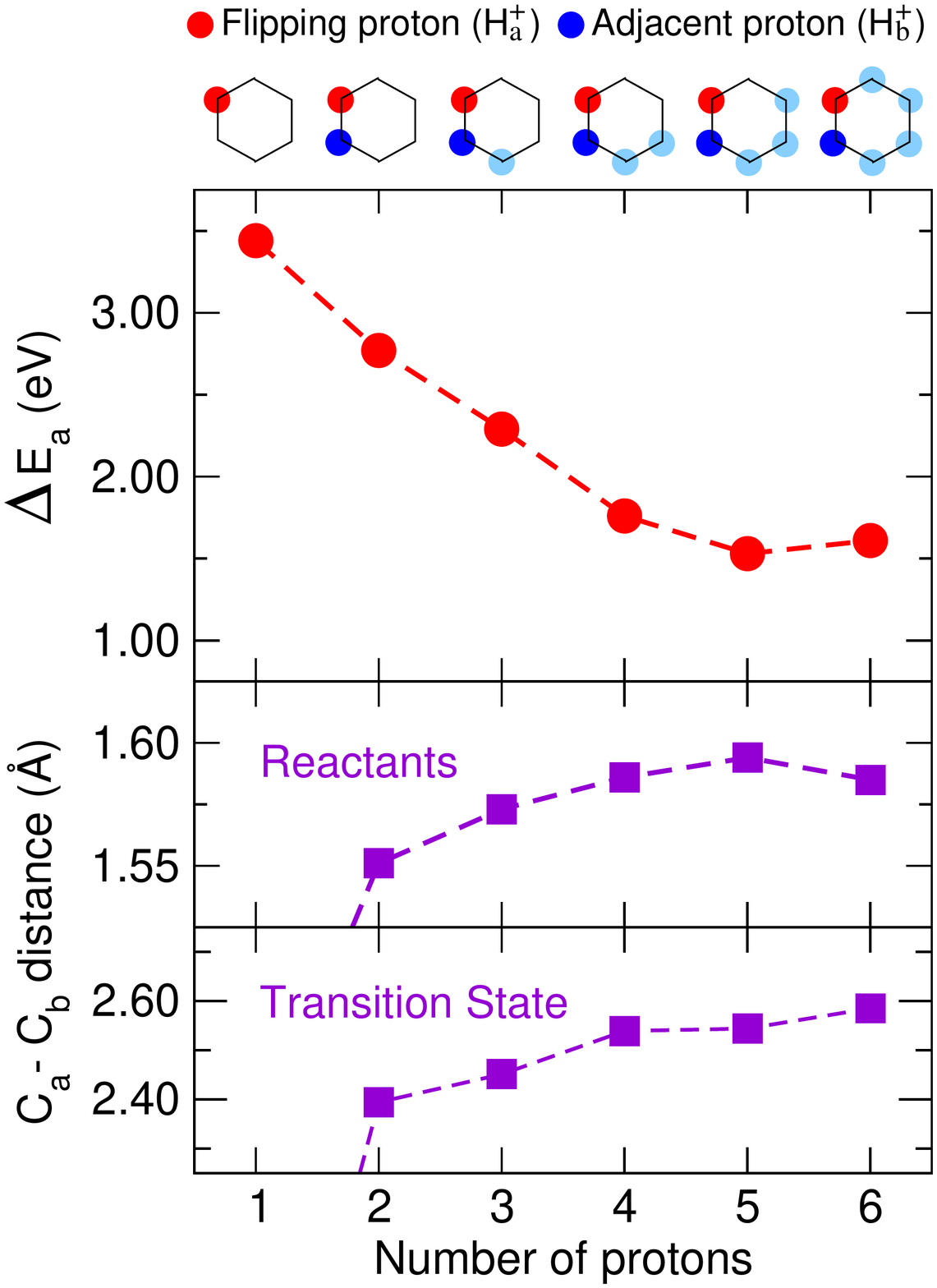}
\caption[]{Upper panel: Activation energy $\Delta E_a$ (in eV) for the permeation of a proton through a
  graphene prototype (circumcoronene) as a function of the number of chemisorbed protons
  along the central carbon ring. Initial configurations of the protons are schematically
  depicted in the upper part of the figure, where red and dark blue circles
  represent flipping (H$_a^+$) and adjacent (H$_b^+$) protons. Middle panel: distances
  between the corresponding carbon atoms C$_a$ and C$_b$ in the reactant state. Lower panel: same as
  middle panel for the transition state.}
\label{fig4}
\end{figure}

Additional adsorbed protons have been consecutively added along the central ring of the
circumcoronene molecule and the results for the proton permeation are collected in
Table ~\ref{tab:1} and in Fig.~\ref{fig4}. Note that the flipping proton is
that located at one end of the row of chemisorbed protons, as indicated schematically in the upper part of
Fig.~\ref{fig4}. It is found that the activation energy further decreases with the protonation
degree, reaching a minimum of $\Delta E_a$=1.53 eV for $n=5$ (less than half of the value for
$n$=1) and then slightly increasing to 1.61 eV for $n=6$. The corresponding enthalpies are
roughly 0.10 eV smaller than the activation energies, in accordance with the finding that
the zero point energy is larger for reactants than for the TS. On the other hand, as seen
in Fig.~\ref{fig4} (middle panel), the C$_a$-C$_b$ distance in the reactants state monotonously
increases with $n$ up to $n=5$. This is a consequence of the amplification of the ring area,
in turn due to the increasing number of single C-C bonds and effects of steric repulsion between hydrogens.
This feature probably facilitates a larger C$_a$-C$_b$ distance in the TS
(as indeed seen in the lower panel of Fig.~\ref{fig4}) and therefore 
an easier insertion of the proton between the two carbon atoms (a lower energy barrier).
In fact, the C$_b$-H$_a^+$ distance increases from 1.32 \AA \, for
$n=$2 to 1.52 \AA \, for $n=$6 whereas the C$_a$-H$_a^+$ distance stays in the range of 1.07-1.08 \AA \,
in the whole $n$ range studied. 
  
>From Table ~\ref{tab:1} it can also be seen that (except for $n$=1 where initial and final states
are equivalent) the flipping process is exothermic with electronic energy and enthalpy variations
($\Delta E_r$ and $\Delta H_r$) ranging between -0.4 to -0.5 eV for $n$=2-5. This result is not
surprising since in related systems as hydrogenated graphene\cite{Pumera-2013} the most stable states
involve adjacent carbon atoms, each one linking hydrogen atoms at opposite sides of the layer.
Moreover, note that $\Delta E_r$ drops to about -1.20 eV for $n=6$, indicating that the permeation
process is globally more favorable when the graphenic ring protonation is complete. In this case a
sp$^3$ configuration of the carbon atom bound to the flipped proton appears to be even more feasible
as it is connected with two carbon atoms supporting protons on the other side, in an arrangement
that is more similar to the typical chair conformation of graphane\cite{Pumera-2013}.

Interestingly, we have found that the permeation process exhibits different activation barriers
and exothermicities depending on the position of the proton within a given row along a carbon
ring. For instance, we have studied the case where the flipping proton is the central one in a
row of five protons  and compare the results with those of the $n=5$ case
reported above, which shares the same reactant state but where the flipping proton is one at
the end of the row (left panel of Fig.~\ref{fig3}). The process occurs through the same kind of
insertion TS  but the activation energy is considerably larger
(2.54 eV vs. 1.53 eV).
We have noticed that at the TS the C$_a$-C$_b$ distance is smaller than in the previous case;
there are also some other differences in size, shape and flatness of the central and adjacent
rings between the two TS being compared. In addition, exothermicity of this process is larger
(-0.83 eV) than for the flipping of the external proton (-0.46 eV). As already discussed for 
$n$=6, the greater stability of the products can be related with a larger facility
of the carbon atom linking the flipped proton to be arranged in a sp$^3$ geometry.

\begin{figure}
\includegraphics[width=6.5cm,angle=0.]{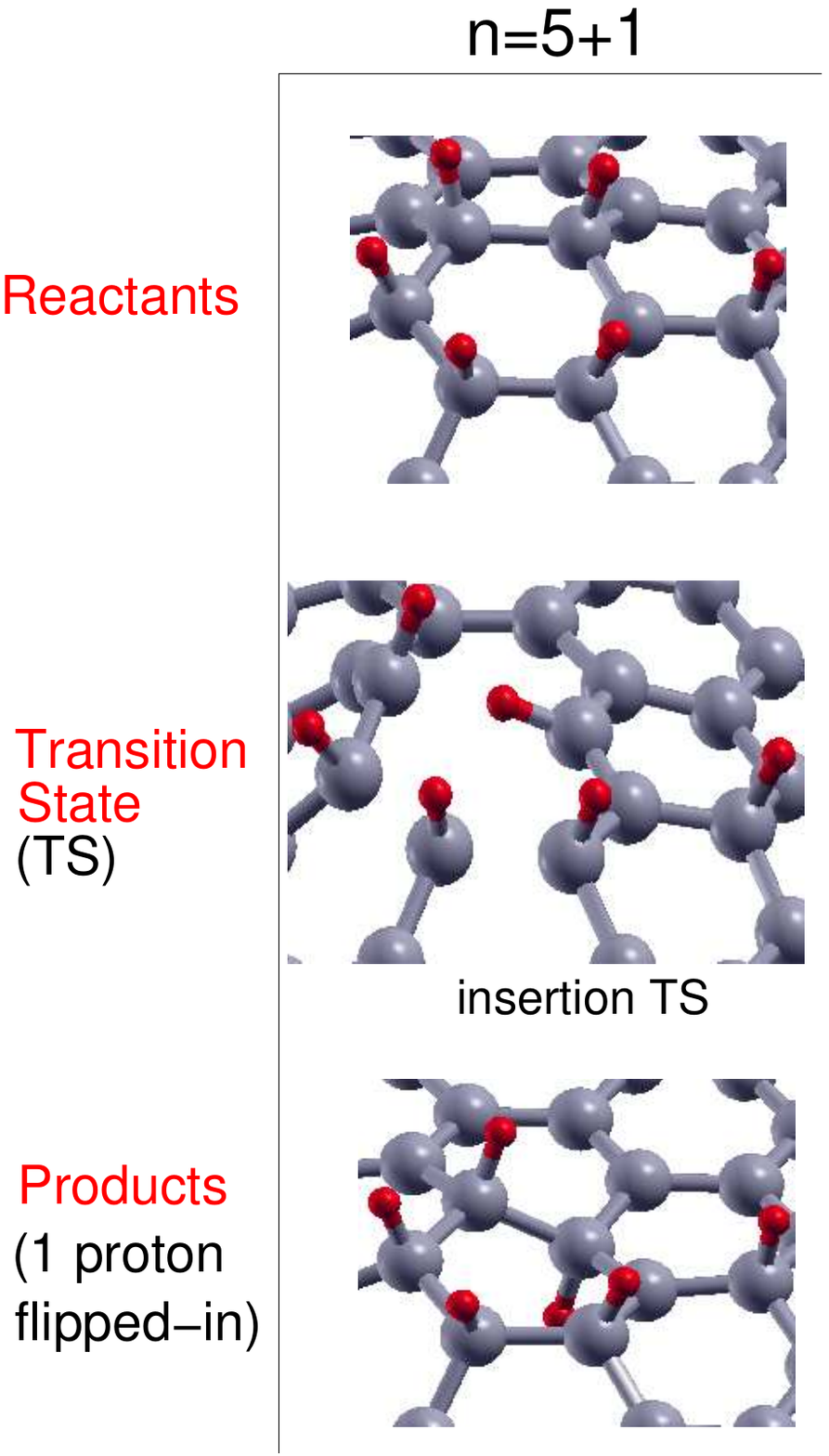}
\caption[]{ Same as in Figs.\ref{fig2} and \ref{fig3} for the case of five protons on a ring plus
  another one on a nearest neighbor ring.}
\label{fig5}
\end{figure}

As can be foreseen from the previous results, it is a real challenge to identify 
the optimum configuration for the proton transport among a large number of different initial
proton distributions. We have analyzed some more 
candidates and in Table ~\ref{tab:1} we show the obtained optimum configuration,
named $5+1$, corresponding to five protons adsorbed on the central ring plus another one on a nearest
neighbor ring, as reported in Fig.~\ref{fig5}. The related activation energy is only 1.0 eV
which already can be considered consistent with the experimental determination\cite{Hu-2014},
especially taking into account that a further reduction when properly including tunneling effects
could be expected. This brings strong support to the newly proposed mechanism
which could provide a new perspective for the explanation of the experimental findings. 

 Moreover, we would also like to address the question of the reliability of
  present findings with respect to the size of the finite graphene prototype. 
To do
  that we have calculated the corresponding activation energy for the proton
  flipping process also for circumcircumcoronene (C$_{96}$H$_{24}$) whose
  estimations.  Even if the energy barriers 
are globally larger
  for the larger prototype a similar trend can be observed in both cases: 
in fact, they decrease
  with the number of chemisorbed protons up to $n$=5 and, more importantly, reach a similar value
around 1 eV for the $n$=5+1 case. These results suggest that low activation energies, 
which are compatible with the experimental findings, are expected also for
larger graphene flakes.

Finally, there have been two previous theoretical studies which deserve special
mention in connection
with the present work,  both of them already briefly mentioned in the Introduction. First we
recall the work of Feng {\em et al}\cite{Feng-2017} where the importance of locally saturating
carbon chemisorption sites via hydrogenation together with the associated change in carbon
hybridization proved crucial in determining lower barriers for proton transport. We would
like to stress some important differences with our proposed new mechanism. First, our model
strictly includes the addition of protons as opposed to hydrogen atoms thus
lending our approach closer to the proton permeation experimental conditions.
Second and most important, the permeation process reported by Feng {\em et al} involves
  a completely different mechanism, as the proton is not initially chemisorbed (as in the
  present case) since the nearby chemisorbed sites are already saturated by hydrogenation, this
  unstability in the initial state leading to an important lowering of the barrier. Moreover,
  proton penetrates through the hole of the carbon ring, in contrast with the insertion
  transition state presented here. In this way, both mechanisms are qualitatively different,
  but with one not excluding the other.

The process reported here does bear a resemblance to the flip-in hydrogen insertion
  mechanism across a (5,5) SWCNT by Lee {\em et al}\cite{Lee-JACS-2001}.  
  By means of periodic
  density-functional tight-binding calculations, they found a low energy path (with
  $\Delta E_a$= 1.51 eV) for hydrogen atom insertion into the middle of a very stretched C-C bond
  (detailed structure of the transition state not provided), a bond that is exothermically
  recovered after the hydrogen atom has flipped-in.
  These similarities were not necessarily foreseeable, given the differences in the structure
  of the (very curved) (5,5) SWCNT and (flat) graphene and the well-known dependence
  of reactivity on the curvature of the carbon substrate\cite{Chen2003}.
However, they note that the flip-in process is efficient only if the nanotube is
completely saturated, while an analogous conclusion for protonated graphene is not suggested from
the present explorations.  Also, the authors report even lower activation 
barriers for the subsequent flipping-in of nearby hydrogen atoms, 
a finding that is opposite to the test carried out with a second proton.

\section{Summary and Outlook}

To summarize, we have reported DFT calculations of the permeation (flipping) of a proton
through multiprotonated graphene, using circumcoronene as its molecular prototype.
A new mechanism involving relatively low energy barriers (down to about 1.0 eV) has been found to occur
when there is at least one other chemisorbed proton next to the flipping one. The
corresponding transition state is characterized by the insertion of the proton into the middle
of a C-C bond and a rearrangement of the hybridizations of the involved carbon atoms, with a
recovering of that bond after the flipping is completed. The nature of this transition
state as well as the enlargement of the C-C distances in the initial state is at the origin
of the reduced activation energy as compared with the flipping of an isolated chemisorbed proton.

The lowest reported energy barrier is close to the experimentally measured activation energy
(0.8 eV), and some contributions not taken into account in the present work -such as
solvent effects\cite{Kroes-2017,XuAo-2017,Ekanayake-2017},
nuclear quantum effects\cite{Ekanayake-2017,Poltavsky-2018,Mazzuca-2018,Feng-2017,heliumisotop:15,Gijon-2017}
and bias potential\cite{Hu-2014,Lozada-2016}- could actually play a role in further decreasing it.
In the context of these experiments and within the frame of the adsorption-penetration model
mentioned in the Introduction, it would be paramount to study the multiprotonation of
graphene by proton transfer from the aqueous environment.
 A preliminary exploration indicates that the energy balance for the proton transfer from
  hydronium ions is strongly dependent on the size of the graphene prototype and that many
  effects such as the formation of adducts and the simulation of the aqueous 
medium should be carefully addressed.
  Moreover, we believe that it is
  worth to investigate whether related systems such as hydrogenated graphene or
hydrogenated/protonated hexagonal boron nitride exhibit permeation processes with mechanisms
 analogous to that here reported. Work along some of these directions is in progress.

\section*{Acknowledgments}

We gratefully thank N. Halberstadt and A. Beswick for valuable directions provided. We also
thank Maciej Gutowski for many useful discussions on extended systems.
The work has been funded by the Spanish grant FIS2017-84391-C2-2-P. 
Allocation of computing time by CESGA (Spain) is also  acknowledged.
RHL would like to thank CONACYT for a sabbatical scholarship.




\end{document}